\documentclass[noshowpacs,pra,twocolumn,english]{revtex4}

\pdfoutput=1

\usepackage{mathtools}
\usepackage[T1]{fontenc}
\usepackage{graphicx}
\usepackage{dcolumn}
\usepackage{bm}
\usepackage{natbib}
\usepackage{amsmath}
\usepackage{here}
\usepackage{color}
\usepackage[normalem]{ulem}
\usepackage{epstopdf}
\usepackage{pdfpages}
\usepackage{siunitx}

\usepackage{amsthm}
\usepackage{enumerate}

\theoremstyle{definition}
\newtheorem{obs}{Observation}
\newtheorem{lemma}[obs]{Lemma}

\usepackage{algorithm}
\usepackage[noend]{algpseudocode}

\begin{document}

\title{Job Shop Scheduling Solver based on Quantum Annealing}

\author{Davide Venturelli$^{1,2}$, {Dominic J.J. Marchand}$^{3}$, {Galo Rojo}$^{3}$}

\affiliation{$^{1}$Quantum Artificial Intelligence Laboratory (QuAIL), NASA Ames\\$^{2}$U.S.R.A. Research Institute for Advanced Computer Science (RIACS)\\$^{3}$1QB Information Technologies (1QBit)}

\begin{abstract}
Quantum annealing is emerging as a promising near-term quantum computing approach
 to solving combinatorial optimization problems. A solver for the job-shop scheduling problem that makes use of a quantum annealer is presented in detail. Inspired by methods used for constraint satisfaction problem (CSP) formulations, we first define the makespan-minimization problem as a series of decision instances before casting each instance into a time-indexed quadratic unconstrained binary
 optimization. Several pre-processing and graph-embedding strategies are employed to compile optimally parametrized families of problems for scheduling instances 
 on the D-Wave Systems' Vesuvius quantum annealer (D-Wave Two). Problem simplifications
 and partitioning algorithms, including variable pruning, are discussed and the results from the processor are compared against classical global-optimum solvers.
\end{abstract}

\maketitle

\section{I. Introduction\label{sec:Introduction}}

The commercialization and independent benchmarking \cite{johnson2010scalable, boixo2014evidence, ronnow2014defining, mcgeoch2013experimental} of
 quantum annealers based on superconducting qubits has sparked
a surge of interest for near-term practical applications of quantum
analog computation in the optimization research community.
Many of the early proposals for running useful problems arising
in space science \cite{smelyanskiy2012near} have
been adapted and have seen small-scale testing on the D-Wave
Two
processor \cite{RieffelVenturelli_qip}. The best procedure
for comparison of quantum analog performance with traditional digital
methods is still under debate \cite{ronnow2014defining, HenProbing2015, newKatzgraber} and remains mostly
speculative due to the limited number of qubits on the currently
available hardware. While waiting for
the technology to scale up to more significant sizes, there is an 
increasing interest in the identification of small problems which are nevertheless 
computationally challenging and useful. One approach in this direction has been 
pursued in \cite{RieffelVenturelli_proc}, and consisted in identifying parametrized 
ensembles of random instances of operational planning problems of increasing sizes
that can be shown to be on the verge of a solvable-unsolvable phase
transition. This condition should be sufficient to observe an asymptotic
exponential scaling of runtimes, even for instances of relatively small
size, potentially testable on current- or next-generation
D-Wave hardware. An empirical takeaway from~\cite{RieffelVenturelli_qip} (validated
also by experimental results in \cite{ogorman2015compiling, Venturelli2015SK}) was that the established
programming and program running techniques for quantum annealers seem
to be particularly amenable to scheduling problems, allowing
for an efficient mapping and good performance compared to other applied
problem classes like automated navigation and Bayesian-network
structure learning \cite{OGorman}.

Motivated by these first results, and with the intention to challenge current
technologies on hard problems of practical value, we herein formulate a
quantum annealing version of the job-shop scheduling problem (JSP). 
The JSP is essentially a general paradigmatic constraint satisfaction problem (CSP) framework for the problem of optimizing
the allocation of resources required for the execution of sequences of
operations with constraints on location and time.
We provide compilation and running strategies for this problem using original and
traditional techniques for parametrizing ensembles of instances.
Results from the D-Wave Two are compared with classical exact
solvers. The JSP has earned a reputation for being especially intractable, a claim supported
by the fact that the best general-purpose solvers (CPLEX, Gurobi Optimizer, SCIP) struggle
with instances as small as 10 machines and 10 jobs (10~x~10)~\cite{kurevisiting}. Indeed, some known
20~x~15 instances often used for benchmarking still have not been solved to
optimality even by the best special-purpose solvers~\cite{jainReview}, and 20~x~20 instances are typically completely intractable.
We note that this early work constitutes a wide-ranging survey of possible techniques and
research directions and leave a more in-depth exploration of these topics for
future work.

\subsection{Problem definition and conventions}
Typically the JSP consists of a set of jobs
$\mathcal{J}=\{{\bf j}_{1},\dots, {\bf j}_{N}\}$
that must be scheduled on a set of machines
$\mathcal{M}=\{{\bf m}_{1},\dots, {\bf m}_{M}\}$.
Each job consists of a sequence of operations that must be performed
in a predefined order $${\bf j}_{n}=\{O_{n1}\to O_{n2}\to\dots\to O_{nL_n}\}.$$
Job ${\bf j}_n$ is assumed to have $L_n$ operations.
Each operation $O_{nj}$ has an integer execution time $p_{nj}$ (a value of zero is allowed) 
 and has to be executed by an assigned machine
${\bf m}_{q_{nj}}\in\mathcal{M}$, where $q_{nj}$ is the index of the assigned machine. There can only be one operation running on any given machine at any given point in time and each operation of a job needs to complete before the following one can start.
The usual objective is to schedule all operations in a valid
sequence while minimizing the makespan (i.e., the completion time of
the last running job), although other objective functions can be used. In what follows, we will denote with $\mathcal{T}$ the minimum possible makespan associated with a given JSP instance.

As defined above, the JSP variant we consider is denoted \mbox{${\bf J}M \big| p_{nj}\!\in\![p_\textrm{min},\dots,p_\textrm{max}] \big| C_\textrm{max} $} in the well-known \mbox{$\alpha|\beta|\gamma$} notation, where $p_\textrm{min}$ and $p_\textrm{max}$ are the smallest and largest execution times allowed, respectively. In this notation, ${\bf J}M$ stands for job-shop type on $M$ machines, and $C_\textrm{max}$ means we are optimizing the makespan. 

For notational convenience, we enumerate the operations in
a lexicographical order in such a way that
\begin{eqnarray}
{\bf j}_{1}&=&\{O_{1}\to\dots\to O_{k_{1}}\},\notag\\
\quad {\bf j}_{2}&=&\{O_{k_{1}+1}\to\dots\to O_{k_{2}}\},\notag\\
&\dots&\notag\\
{\bf j}_{N}&=&\{O_{k_{N-1}+1}\to\dots\to O_{k_{N}}\}.\label{eq:Jlexicographic}
\end{eqnarray}
Given the running index over all operations $i\in\{1,\dots, k_{N}\}$,
we let $q_{i}$ be the index of the machine ${\bf m}_{q_{i}}$ responsible
for executing operation $O_{i}$. We define $I_{m}$ to be the set of indices of all of the operations that have to be executed
on machine ${\bf m}_{m}$, i.e., $I_{m}=\{i: q_{i}=m\}$. The execution
time of operation $O_{i}$ is now simply denoted $p_{i}$. 

A priori, a job can use the same machine more than once, or use only a fraction of the $M$ available machines. For benchmarking purposes, it is customary to restrict a study to the problems of a specific family. In this work, we define a ratio $\theta$ that specifies the fraction of the total number of machines that is used by each job, assuming no repetition when $\theta\le 1$. For example, a ratio of 0.5 means that each job uses only $0.5M$ distinct machines. 

\subsection{Quantum annealing formulation}
In this work, we seek a suitable formulation of the JSP for a quantum annealing optimizer (such as the D-Wave Two). The optimizer is best described as an oracle that solves an Ising problem with a given probability~\cite{boros2001pseudoboolean}. This Ising problem is equivalent to a quadratic unconstrained binary optimization (QUBO) problem~\cite{ogorman2015compiling}. The binary polynomial associated with a QUBO problem can be depicted as a graph, with nodes representing variables and values attached to nodes and edges representing linear and quadratic terms, respectively. The QUBO solver can similarly be represented as a graph where nodes represents qubits and edges represent the allowed connectivity. The optimizer is expected to find the global minimum with some probability which itself depends on the problem and the device's parameters. The device is not an ideal oracle: its limitations, with regard to precision, connectivity, and number of variables, must be considered to achieve the best possible results. 
As is customary, we rely on the classical procedure known as embedding to adapt the connectivity of the solver to the problem at hand. This procedure is described in a number of quantum annealing papers \cite{RieffelVenturelli_qip, Venturelli2015SK}. During this procedure, two or more variables can be forced to take on the same value by including additional constraints in the model. In the underlying Ising model, this is achieved by introducing a large ferromagnetic (negative) coupling $J_{\textrm{F}}$ between two spins. The embedding process modifies the QUBO problem accordingly and one should not confuse the \emph{logical} QUBO problem value, which depends on the QUBO problem and the state considered, with the Ising problem energy seen by the optimizer (which additionally depends on the extra constraints and the solver's parameters, such as $J_{\textrm{F}}$).

We distinguish between the \emph{optimization} version of the JSP, in which we seek a valid schedule with a minimal makespan, and the \emph{decision} version, which is limited to validating whether or not a solution exists with a makespan smaller than or equal to a user-specified timespan $T$. We focus exclusively on the decision version and later describe how to implement a full optimization version based on a binary search. We note that the decision formulation where jobs are constrained to fixed time windows is sometimes referred in the literature as the job-shop CSP formulation \cite{cheng1997applying, garrido2000heuristic}, and our study will refer to those instances where the jobs share a common deadline $T$.

\section{II. QUBO problem formulation\label{sec:JSP-and-its}}

While there are several ways the JSP can be formulated, such as the rank-based formulation \cite{Wagner} or the disjunctive formulation \cite{Manne}, 
our formulation is based on a straightforward time-indexed representation particularly amenable to quantum annealers 
(a comparative study of mappings for planning and scheduling problems can be found in \cite{ogorman2015compiling}). We assign a set of binary variables for each operation, corresponding to the various possible discrete starting times the operation can have:
\begin{equation}
x_{i,t}=\left\{ \begin{array}{ll}
1 & :\text{ operation }O_{i}\text{ starts at time }t,\\
0 & :\text{ otherwise}.
\end{array}\right.\label{eq:bit definition}
\end{equation}
Here $t$ is bounded from above by the timespan $T$, which represents
the maximum time we allow for the jobs to complete. The timespan itself is bounded from above by the total work of the problem, that is, the sum of the execution times of all operations.

\begin{figure}[h!]
\begin{centering}
\includegraphics[scale=0.42]{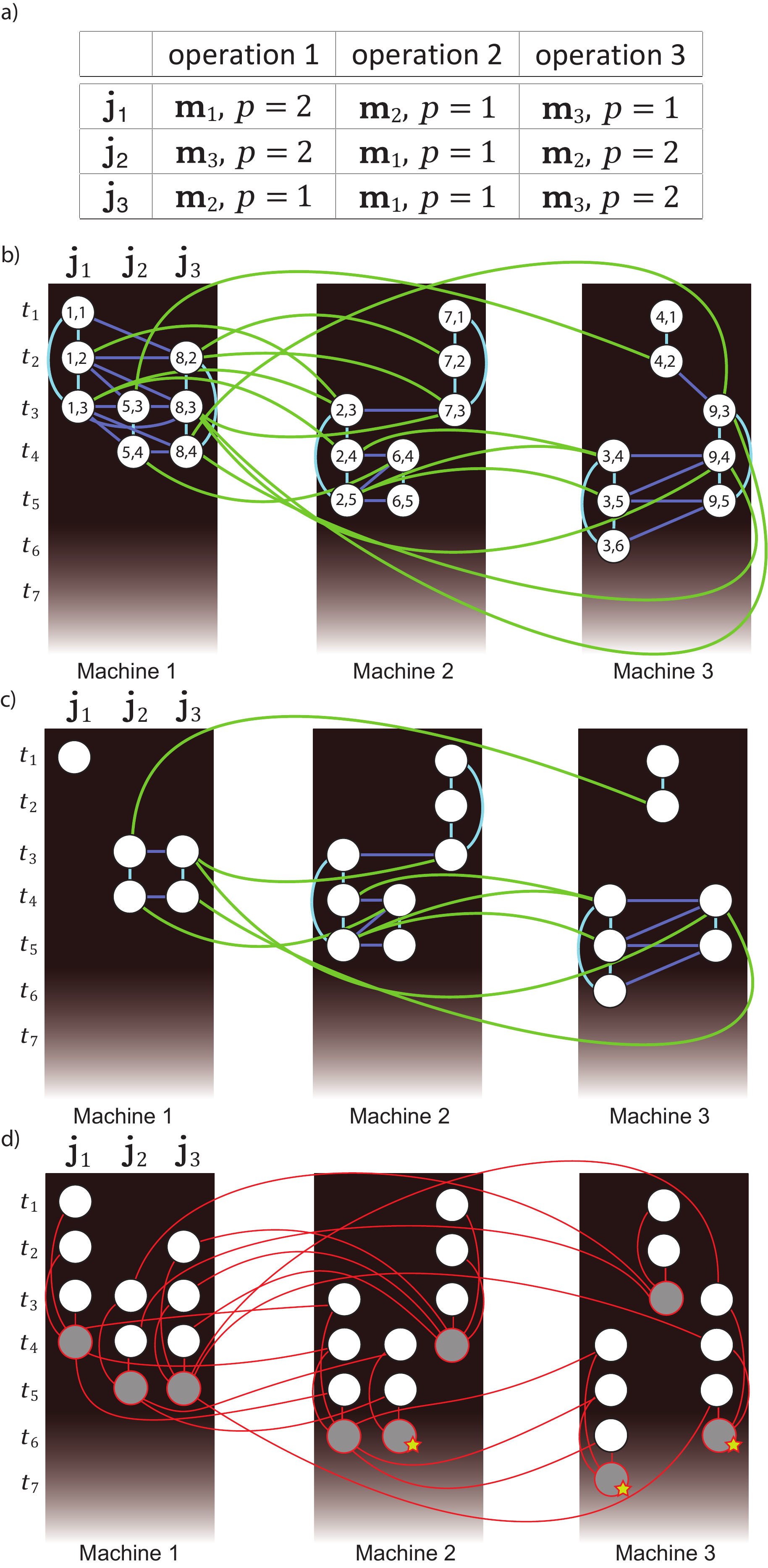}
\par\end{centering}

\caption{a) Table representation of an example 3~x~3 instance whose execution
times have been randomly selected to be either 1 or 2 time units.
b) Pictorial view of the QUBO mapping of the above example for $H_{T = 6}$.
Green, purple, and cyan edges refer respectively to $h_{1}$, $h_{2}$,
and $h_{3}$ quadratic coupling terms (Eqs. \ref{eq:h1}--\ref{eq:h3}).
Each circle represents a bit with its $i,t$ index as in Eq. \ref{eq:bit definition}.
c) The same QUBO problem as in (b) after the variable pruning procedure detailed in the section on QUBO formulation refinements. Isolated qubits are bits with fixed assignments that can be eliminated from the final QUBO problem.
d) The same QUBO problem as in (b) for $H_{T = 7}$. Previously displayed edges in the
above figure are omitted. Red edges/circles represent the variations
with respect to $H_{T = 6}$. Yellow stars indicate the bits which are
penalized with local fields for timespan discrimination.}
\label{fig:QUBOfigure}
\end{figure}

\subsection{Constraints}

We account for the various constraints by adding penalty terms to the QUBO problem.
For example, an operation must start once and only once, leading to the constraint
and associated penalty function
\begin{equation}
\left(\sum_{t}x_{i,t}=1 \text{ for each }i\right)\to\sum_{i}\left(\sum_{t}x_{i,t}-1\right)^{2}.\label{start_once}
\end{equation}
There can only be one job running on each machine at any given
point in time, which expressed as quadratic constraints yields
\begin{equation}
\sum_{(i,t,k,t')\in R_{m}}x_{i,t}x_{k,t'}=0 \text{ for each }m,
\label{no_collisions}
\end{equation}
where $R_{m}=A_{m}\cup B_{m}$ and
\begin{eqnarray*}
A_{m}&=&\{(i,t,k,t^{\prime}):(i,k)\in I_{m}\times I_{m},\\
&&i\neq k,0\leq t,t^{\prime}\leq T,0<t^{\prime}-t<p_{i}\},\\
B_{m}&=&\{(i,t,k,t^{\prime}):(i,k)\in I_{m}\times I_{m},\\
&&i<k,t^{\prime}=t,p_{i}>0,p_{j}>0 \}.
\end{eqnarray*}
The set $A_{m}$ is defined so that the constraint forbids operation
$O_{j}$ from starting at $t^{\prime}$ if there is another
operation $O_{i}$ still running, which happens if $O_{i}$ started
at time $t$ and $t^{\prime}-t$ is less than $p_{i}$. The set
$B_{m}$ is defined so that two jobs cannot start at the same time,
unless at least one of them has an execution time equal to zero.
Finally, the order of the operations within a job are enforced with
\begin{equation}
\sum_{\substack{k_{n-1}< i<k_{n}\\
t+p_{i}>t'}
}x_{i,t}x_{i+1,t^{\prime}}\quad\text{ for each }n,\label{precedence}
\end{equation}
which counts the number of precedence violations between consecutive
operations only.

The resulting classical objective function (Hamiltonian) is given by
\begin{equation}
H_{T}(\bar{x})=\eta h_{1}(\bar{x})+\alpha h_{\text{2}}(\bar{x})+\beta h_{\text{3}}(\bar{x}),
\label{eq:objective_function}
\end{equation}
where
\begin{eqnarray}
h_{1}(\bar{x}) & = & \sum_{n}\left(\sum_{\substack{k_{n-1} < i < k_{n}\\
t+p_{i}>t'
}
}x_{i,t}x_{i+1,t^{\prime}}\right),\label{eq:h1}\\
h_{\text{2}}(\bar{x}) & = & \sum_{m}\left(\sum_{(i,t,k,t^{\prime})\in R_{m}}x_{i,t}x_{k,t^{\prime}}\right),\label{eq:h2}\\
h_{\text{3}}(\bar{x}) & = & \sum_{i}\left(\sum_{t}x_{i,t}-1\right)^{2},\label{eq:h3}
\end{eqnarray}
and the penalty constants $\eta$, $\alpha$, and $\beta$ are required to be larger than $0$
to ensure that unfeasible solutions do not have a lower energy than
the ground state(s). As expected for a decision problem, we note that the minimum of
$H_{T}$ is $0$ and it is
only reached if a schedule satisfies all of the constraints.
The index of $H_{T}$ explicitly shows the dependence of the Hamiltonian on the timespan
$T$, which affects the number of variables involved. \mbox{Figure~\ref{fig:QUBOfigure}-b}
illustrates the QUBO problem mapping for $H_{T=6}$ for a particular 3~x~3
example \mbox{(Figure~\ref{fig:QUBOfigure}-a)}.

\subsection{Simple variable pruning}
\mbox{Figure \ref{fig:QUBOfigure}-b} also reveals that a significant number of the $N M T$
binary variables required for the mapping can be pruned by applying
simple restrictions on the time index $t$ (whose computation is 
polynomial as the system size increases and therefore trivial here). Namely, we can define an
effective release time for each operation corresponding to the sum
of the execution times of the preceding operations in the same job.
A similar upper bound corresponding to the timespan minus all of the
execution times of the subsequent operations of the same job can
be set. The bits corresponding to these invalid starting times can
be eliminated from the QUBO problem altogether since any valid solution would
require them to be strictly zero. This simplification eliminates
an estimated number of variables equal to
$N M\left(M\left\langle p\right\rangle-1\right) $,
where $\left\langle p\right\rangle $ represents the
average execution time of the operations.
This result can be generalized to consider the previously defined ratio $\theta$, such that the total number of variables required after this simple QUBO problem pre-processing is $\theta NM [T - \theta M\langle p\rangle + 1]$.

\section{III. QUBO formulation refinements}

Although the above formulation proves sufficient for running JSPs on the D-Wave machine, we explore a few potential refinements. The first pushes the limit of simple variable pruning by considering more advanced criteria for reducing the possible execution window of each task. The second refinement proposes a compromise between the decision version of the JSP and a full optimization version.

\subsection{Window shaving}

In the time-index formalism, reducing the execution windows of operations (i.e., shaving) \cite{Martin}, or in the disjunctive approach, adjusting the \emph{heads} and \emph{tails} of operations \cite{Carlier, Peridy}, or more generally, by applying constraints propagation techniques (e.g. \cite{caseau}), together constitute the basis for a number of classical approaches to solving the JSP. Shaving is sometimes used as a pre-processing step or as a way to obtain a lower bound on the makespan before applying other methods. The interest from our perspective is to showcase how such classical techniques remain relevant, without straying from our quantum annealing approach, when applied to the problem of pruning as many variables as possible. This enables larger problems to be considered and improves the success rate of embeddability in general (see Figure \ref{fig:Embedding}), without significantly affecting the order of magnitude of the overall time to solution in the asymptotic regime. Further immediate advantages of reducing the required number of qubits become apparent during the compilation of JSP instances for the D-Wave device due to the associated embedding overhead that depends directly on the number of variables. The shaving process is typically handled by a classical algorithm whose worst-case complexity remains polynomial. While this does not negatively impact  the fundamental complexity of solving JSP instances, for \emph{pragmatic} benchmarking the execution time needs to be taken into account and added to the quantum annealing runtime to properly report the time to solution of the whole algorithm.

Different elimination rules can be applied. We focus herein on the iterated Carlier and Pinson (ICP) procedure \cite{Carlier} reviewed in the appendix with worst-case complexity given by $\mathcal{O}(N^2M^2T\log(N))$. Instead of looking at the one-job sub-problems and their constraints to eliminate variables, as we did for the simple pruning, we look at the one-machine sub-problems and their associated constraints to further prune variables. An example of the resulting QUBO problem is presented in \mbox{Figure~\ref{fig:QUBOfigure}-c}.

\subsection{Timespan discrimination}

We explore a method of extracting more information regarding the actual optimal makespan of a problem within a single call to the solver by breaking the degeneracy of the ground states and spreading them over some finite energy scale, distinguishing the energy of valid schedules on the basis of their makespan. Taken to the extreme, this approach would amount to solving the full optimization problem. We find that the resulting QUBO problem is poorly suited to a solver with limited precision, so a balance must be struck between extra information and the precision requirement. A systematic study of how best to balance the amount of information obtained versus the extra cost will be the subject of future work.

We propose to add a number of linear terms, or local fields, to the QUBO problem to slightly penalize valid solutions with larger makespans. We do this by adding a cost to the last operation of each job, that is, the set $\{O_{k_1},\dots,O_{k_N}\}$. At the same time, we require that the new range of energy over which the feasible solutions are spread stays within the minimum logical QUBO problem's gap given by $\Delta E=\min\{\eta, \alpha, \beta\}$. If the solver's precision can accomodate $K$ distinguishable energy bins, then makespans within $[T-K,\,T]$ can be immediately identified from their energy values. The procedure is illustrated in \mbox{Figure \ref{fig:QUBOfigure}-d} and some implications are discussed in the appendix.

\section{IV. Ensemble pre-characterization and compilation}\label{section_charac_compil}

We now turn to a few important elements of our computational strategy for solving JSP instances. We first show how a careful pre-characterization of classes of random JSP instances, representative of the problems to be run on the quantum optimizer, provides very useful information regarding the shape of the search space for $\mathcal{T}$. We then describe how instances are compiled to run on the actual hardware.

\subsection{Makespan Estimation}

\begin{figure}[h!]
\begin{centering}
\includegraphics[scale=0.36]{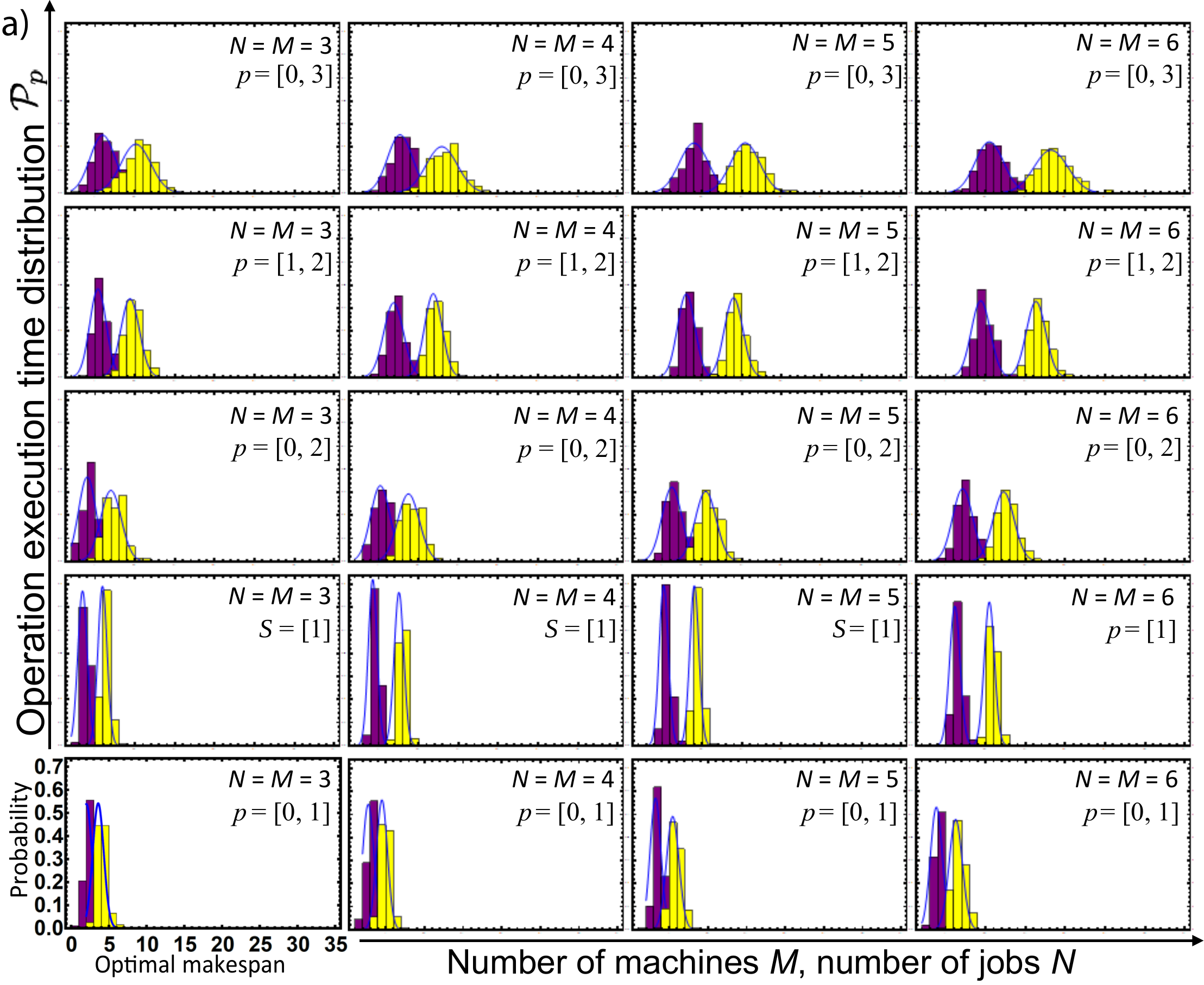}
\par\end{centering}

\caption{a) Normalized histograms of optimal makespans $\mathcal{T}$ for parametrized families
of JSP instances with $N=M$, $\mathcal{P}_{p}$ on the y-axis, $\theta=1$ (yellow), and $\theta=0.5$ (purple). The distributions
are histograms of occurrences for 1000 random instances, fitted with a Gaussian function of mean $\langle\mathcal{T}\rangle$. We note that the width of the distributions increases as the range of the execution times $\mathcal{P}_{p}$ increases, for fixed $\langle p \rangle$. The mean and the variance are well fitted respectively by $\langle \mathcal{T}\rangle=A_{\mathcal{T}}Np_{\textrm{min}}+B_{\mathcal{T}}Np_{\textrm{max}}$ and $\sigma=\sigma_0+C_{\sigma}\langle \mathcal{T}\rangle+A_{\sigma}p_{\textrm{min}}+B_{\sigma}p_{\textrm{max}}$, with $A_{\mathcal{T}}=0.67$, $B_{\mathcal{T}}=0.82$, $\sigma_0=0.7$, $A_{\sigma}=-0.03$,  $B_{\sigma}=0.43$, and $C_{\sigma}=0.003$.}
\label{fig:precharacterization}
\end{figure}

In Figure~\ref{fig:precharacterization}, we show the distributions of the optimal makespans $\mathcal{T}$ for different ensembles of instances parametrized by their size $N=M$, by the possible values of task durations $\mathcal{P}_{p}=\{p_{\textrm{min}},\dots,p_{\textrm{max}}\}$, and by the ratio $\theta\le1$ of the number of machines used by each job. Instances are generated randomly by selecting $\theta M$ distinct machines for each job and assigning an execution time to each operation randomly. For each set of parameters, we can compute solutions with a classical exhaustive solver in order to identify the median of the distribution $\langle \mathcal{T}\rangle(N,\mathcal{P}_p,\theta)$ as well as the other quantiles. These could also be inferred from previously solved instances with the proposed annealing solver. The resulting information can be used to guide the binary search required to solve the optimization problem. \mbox{Figure~\ref{fig:precharacterization}} indicates that a normal distribution is an adequate approximation, so we need only to estimate its average $\langle \mathcal{T} \rangle$ and variance $\sigma^2$. Interestingly, from the characterization of the families of instances up to $N=10$ we find that, at least in the region explored, the average minimum makespan $\langle \mathcal{T} \rangle$ is proportional to the average execution time of a job $\langle p \rangle \theta N$, where $\langle p \rangle$ is the mean of $\mathcal{P}_{p}$. 
This linear ansatz allows for the extrapolation of approximate resource requirements for classes of problems which have not yet been pre-characterized, and it constitutes an educated guess for classes of problems which cannot be pre-characterized due to their difficulty or size. The accuracy of these functional forms was verified by computing the relative error of the prediction versus the fit of the makespan distribution of each parametrized family up to $N=M=9$ and $p_{\textrm{max}}=20$ using 200 instances to compute the makespan histogram. The prediction for $\langle \mathcal{T} \rangle$ results are consistently at least 95\% accurate, while the one for $\sigma$ has at worst a 30\% error margin, a very approximate but sufficient model for the current purpose of guiding the binary search.

\subsection{Compilation}

\begin{figure}[h!]
\begin{centering}
\includegraphics[scale=0.6]{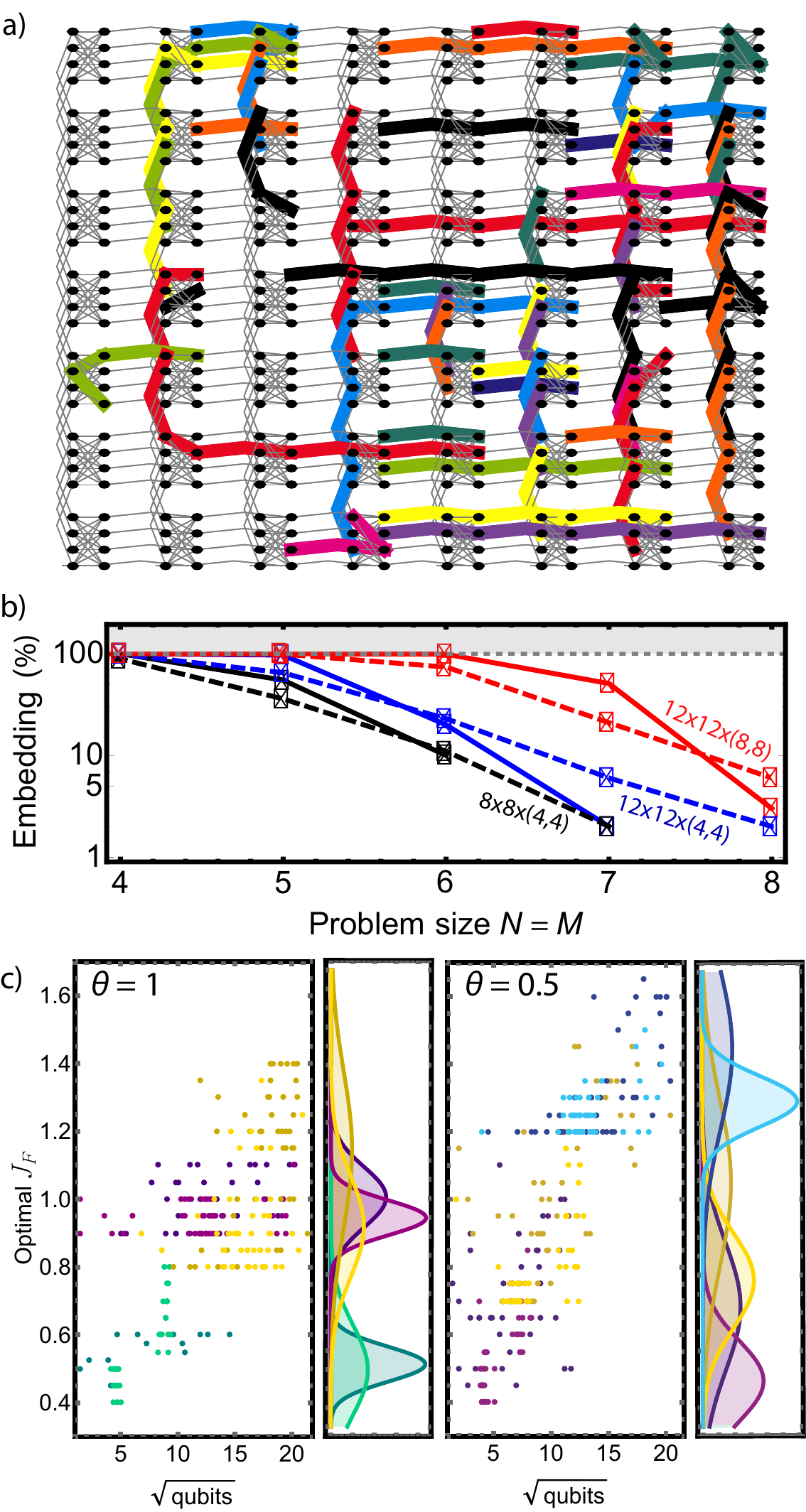}
\par\end{centering}

\caption{a) Example of an embedded JSP instance on NASA's D-Wave Two chip. Each chain of qubits is colored to represent a logical binary variable determined by the embedding. For clarity, active connections between the qubits are not shown. b) Embedding probability as a function of $N=M$ for $\theta=1$ (similar results are observed for $\theta=0.5$). Solid lines refer to $\mathcal{P}_{p}=[1,1]$ and dashed lines refer to $\mathcal{P}_{p}=[0,2]$. 1000 random instances have been generated for each point, and a cutoff of 2 minutes has been set for the heuristic algorithm to find a valid topological embedding. Results for different sizes of Chimera are shown. c)~Optimal parameter-setting analysis for the ensembles of JSP instances we studied. Each point corresponds to the number of qubits and the optimal $J_{F}$ (see main text) of a random instance, and each color represents a parametrized ensemble (green: 3~x~3, purple: 4~x~4, yellow: 5~x~5, blue: 6~x~6; darker colors represent ensembles with $\mathcal{P}_{p}=[1,1]$ as opposed to lighter colors which indicate $\mathcal{P}_{p}=[0,2]$). Distributions on the right of scatter plots represent Gaussian fits of the histogram of the optimal $J_{F}$ for each ensemble. Runtime results are averaged over an ungauged run and 4 additional runs with random gauges~\cite{Perdomo2015}.}

\label{fig:Embedding}
\end{figure}

The graph-minor embedding technique (abbreviated simply ``embedding'') represents the de facto method of recasting the Ising problems to a form compatible with the layout of the annealer's architecture~\cite{LloydEmbedding, ChoiEmbedding}, which for
the D-Wave Two is a Chimera graph~\cite{johnson2010scalable}. Formally, we seek an isomorphism between the problem's QUBO graph and a graph minor of the solver. This procedure has become a standard in solving  applied problems using quantum annealing \cite{RieffelVenturelli_qip, Venturelli2015SK} and can be thought of as the analogue of compilation in a digital computer programming framework during which variables are assigned to hardware registers and memory locations. This process is covered in more details in the appendix. An example of embedding for a 5~x~5 JSP instance with $\theta = 1$ and $T = 7$ is shown in \mbox{Figure~\ref{fig:Embedding}-a}, where the 72 logical variables of the QUBO problem are embedded using 257 qubits of the Chimera graph. Finding the optimal tiling that uses the fewest qubits is NP-hard~\cite{EmbeddingNPhard}, and the standard approach is to employ heuristic algorithms~\cite{DWaveEmbedding}.  In general, for the embedding of time-indexed mixed-integer programming QUBO problems of size $N$ into a graph of degree $~k$, one should expect a quadratic overhead in the number of binary variables on the order of $a N^2$, with $a \le (k-2)^{-1}$ depending on the embedding algorithm and the hardware connectivity~\cite{Venturelli2015SK}. This quadratic scaling is apparent in \mbox{Figure~\ref{fig:Embedding}-b} where we report on the compilation attempts using the algorithm in ~\cite{DWaveEmbedding}. Results are presented for the D-Wave chip installed at NASA Ames at the time of this study, for a larger chip with the same size of Chimera block and connectivity pattern (like the latest chip currently being manufactured by D-Wave Systems), and for a speculative yet-larger chip where the Chimera block is twice as large. We deem a JSP instance embeddable when the respective $H_{T=\mathcal{T}}$  is embeddable, so the decrease in probability of embedding with increasing system size is closely related to the shift and spreading of the optimal makespan distributions for ensembles of increasing size (see Figure \ref{fig:precharacterization}). What we observe is that, with the available algorithms, the current architecture admits embedded JSP instances whose total execution time $N  M \theta \langle p \rangle$ is around 20 time units, while near-future (we estimate 2 years) D-Wave chip architectures could potentially double that. As noted in similar studies~(e.g., \cite{RieffelVenturelli_qip}), graph connectivity has a much more dramatic impact on embeddability than qubit count. 

Once the topological aspect of embedding has been solved, we set the ferromagnetic interactions needed to adapt the connectivity of the solver to the problem at hand. For the purpose of this work, this should be regarded as a technicality necessary to tune the performance of the experimental analog device and we include the results for completeness. Introductory details about the procedure can be found in~\cite{RieffelVenturelli_qip, Venturelli2015SK}. 
 In \mbox{Figure~\ref{fig:Embedding}-c} we show a characterization of the ensemble of JSP instances (parametrized by $N$, $M$, $\theta$, and $\mathcal{P}_p$, as described at the beginning of this section). We present the best ferromagnetic couplings found by runs on the D-Wave machine under the simplification of a uniform ferromagnetic coupling by solving the embedded problems with values of $J_F$ from 0.4 to 1.8 in relative energy units of the largest coupling of the original Ising problem. The run parameters used to determine the best $J_F$ are the same as we report in the following sections, and the problem sets tested correspond to Hamiltonians whose timespan is equal to the sought makespan $H_{T=\mathcal{T}}$. This parameter-setting approach is similar to the one followed in \cite{RieffelVenturelli_qip} for operational planning problems, where the instance ensembles were classified by problem size before compilation. 
What emerges from this preliminary analysis is that each parametrized ensemble can be associated to a distribution of optimal $J_F$ that can be quite wide, especially for the ensembles with $p_{\textrm{min}}=0$ and large $p_{\textrm{max}}$. This spread might discourage the use of the mean value of such a distribution as a predictor of the best $J_F$ to use for the embedding of new untested instances. However, the results from this predictor appear to be better than the more intuitive prediction obtained by correlating the number of qubits after compilation with the optimal $J_F$. This means that for the D-Wave machine to achieve optimal performance on structured problems, it seems to be beneficial to use the information contained in the structure of the logical problem to determine the best parameters. We note that this ``offline'' parameter-setting could be used in combination with ``online'' performance estimation methods such as the ones described in \cite{Perdomo2015} in order to reach the best possible instance-specific $J_F$ with a series of quick experimental runs.  The application of these techniques, together with the testing of alternative offline predictors, will be the subject of future work.

\section{V. Results of test runs and discussion}

A complete quantum annealing JSP solver designed to solve an instance to optimality using our proposed formulation will require the independent solution of several embedded instances $\{H_{T}\}$, each corresponding to a different timespan $T$. Assuming that the embedding time, the machine setup time, and the latency between subsequent operations can all be neglected, due to their being non-fundamental, the running time $\mathtt{T}$ of the approach for a specific JSP instance reduces to the expected {\it{total}} annealing time necessary to find the optimal solution of each embedded instance with a specified minimum target probability $\simeq1$. The probability of ending the annealing cycle in a desired ground state depends, in an essentially unknown way, on the embedded Ising Hamiltonian spectrum, the relaxation properties of the environment, the effect of noise, and the annealing profile. Understanding through an ab initio approach what is the best computational strategy appears to be a formidable undertaking that would require theoretical breakthroughs in the understanding of open-system quantum annealing~\cite{boixo2014computational, smelyanskiychain}, as well as a tailored algorithmic analysis that could take advantage of the problem structure that the annealer needs to solve. For the time being, and for the purposes of this work, it seems much more practical to limit these early investigations to the most relevant instances, and to lay out empirical procedures that work under some general assumptions. More specifically, we focus on solving the CSP version of JSP, not the full optimization problem, and we therefore only benchmark the Hamiltonians with $\mathcal{T}=T$ with the D-Wave machine. We note however that a full optimization solver can be realized by leveraging data analysis of past results on parametrized ensembles and by implementing an adaptive binary search. Full details can be found in a longer version of this work \cite{jsp_arxiv_version}.

\begin{figure}[t!]
\begin{centering}
\includegraphics[scale=0.38]{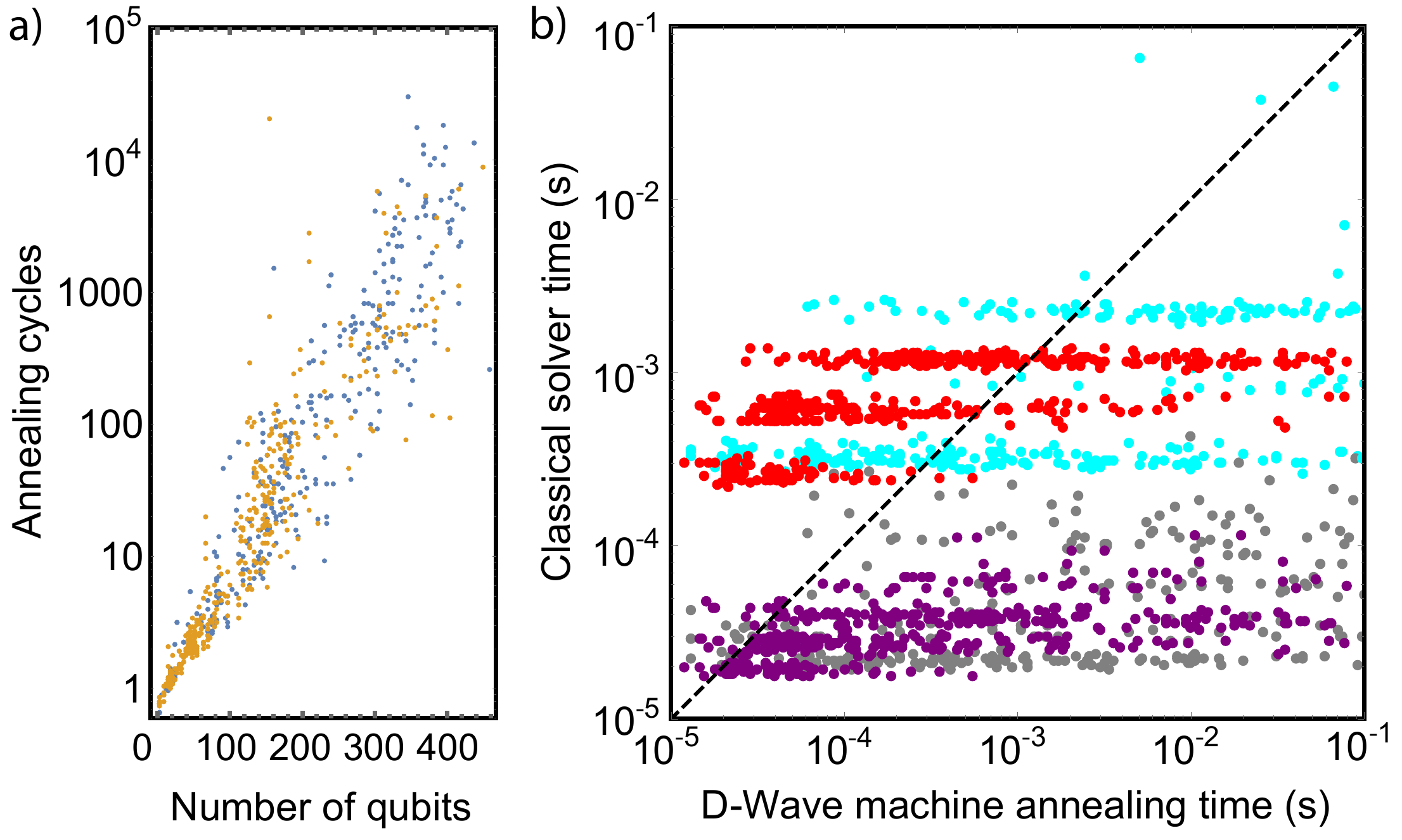}
\par\end{centering}

\caption{{a) Number of repetitions required to solve $H_{\mathcal{T}}$ with the D-Wave Two with a 99\% probability of success (see the appendix). The blue points indicate instances with $\theta=1$ and yellow points correspond to $\theta=0.5$ (they are the same instances and runtimes used for \mbox{Figure~\ref{fig:Embedding}-c}). The number of qubits on the x-axis represents the qubits used after embedding. b) Correlation plot between classical solvers and the D-Wave optimizer. Gray and violet points represent runtimes compared with algorithm {\tt B}, and cyan and red are compared to the {\tt MS} algorithm, respectively, with $\theta=1$ and $\theta=0.5$. All results presented correspond to the best out of 5 gauges selected randomly for every instance. In case the machine returns embedding components whose values are discordant, we apply a majority voting rule to recover a solution within the logical subspace~\cite{Venturelli2015SK, RieffelVenturelli_qip, Perdomo2015, McGeochKing2015, LidarCorrection}. We observe a deviation of about an order of magnitude on the annealing time if we average over 5 gauges instead of picking the best one, indicating that there is considerable room for improvement if we were to apply more-advanced calibration techniques~\cite{perdomoDWtuning}.}}

\label{fig:last}
\end{figure}

On the quantum annealer installed at NASA Ames (it has 509 working qubits; details are presented in~\cite{perdomoDWtuning}), we run hundreds of instances, sampling the ensembles $N=M$ $\in \{3,4,5,6\}$, $\theta$ $\in \{0.5,1\}$, and $P_{p}$ $\in \{ [1,1], [0,2] \}$. For each instance, we report results, such as runtimes, at the most optimal $J_{F}$ among those tested, assuming the application of an optimized parameter-setting procedure along the lines of that described in the previous section. Figure~\ref{fig:last}-a displays the total annealing repetitions required to achieve a 99\% probability of success on the ground state of $H_\mathcal{T}$, with each repetition lasting $t_A=20$~\SI{}{\micro\second}, as a function of the number of qubits in the embedded (and pruned) Hamiltonian. We observe an exponential increase in complexity with increasing Hamiltonian size, for both classes of problems studied. This likely means that while the problems tested are small, the analog optimization procedure intrinsic to the D-Wave device's operation is already subject to the fundamental complexity bottlenecks of the JSP. It is, however, premature to draw conclusions about performance scaling of the technology given the current constraints on calibration procedures, annealing time, etc. Many of these problems are expected to be either overcome or nearly so with the next generation of D-Wave chip, at which point more extensive experimentation will be warranted.

In \mbox{Figure~\ref{fig:last}-b}, we compare the performance of the \mbox{D-Wave} device to two exhaustive classical algorithms in order to gain insight on how current quantum annealing technology compares with paradigmatic classical optimization methods. Leaving the performance of approximate solutions for future work, we chose not to explore the plethora of possible heuristic methods as we operate the D-Wave machine, seeking the global optimum.

The first algorithm,  {\tt B}, detailed in \cite{Brucker}, exploits the disjunctive graph representation and a branch-and-bound strategy that very effectively combines a branching scheme based on selecting the direction of a single disjunctive edge (according to some single-machine constraints), and a technique introduced in \cite{Carlier90} for fixing additional disjunctions (based on a preemptive relaxation). 
It has publicly available code and is considered a well-performing complete solver for the small instances currently accessible to us, while remaining competitive for larger ones even if other classical approaches become more favorable \cite{Beck2010}. {\tt B} has been used in \cite{PhaseTransJSP} to discuss the possibility of a phase transition in the JSP, demonstrating that the random instances with $N=M$ are particularly hard families of problems, not unlike what is observed for the quantum annealing implementation of planning problems based on graph vertex coloring \cite{RieffelVenturelli_proc}.

The second algorithm, {\tt MS}, introduced in \cite{Martin}, proposes a time-based branching scheme where a decision is made at each node to either schedule or delay one of the available operations at the current time. The authors then rely on a series of shaving procedures such as those proposed by \cite{Carlier} to determine the new bound and whether the choice leads to valid schedules. This algorithm is a natural comparison with the present quantum annealing approach as it solves the decision version of the JSP in a very similar fashion to the time-indexed formulation we have implemented on the D-Wave machine, and it makes use of the same shaving technique that we adapted as a pre-processing step for variable pruning. However, we should mention that the variable pruning that we implemented to simplify $H_T$ is employed at each node of the classical branch and bound algorithm, so the overall computational time of {\tt MS} is usually much more important than our one-pass pre-processing step, and in general its runtime does not scale polynomially with the problem size.

What is apparent from the correlation plot in \mbox{Figure~\ref{fig:last}-b} is that the D-Wave machine is easily outperformed by a classical algorithm run on a modern single-core processor, and that the problem sizes tested in this study are still too small for the asymptotic behavior of the classical algorithms to be clearly demonstrated and measured.
The comparison between the D-Wave machine's solution time for $H_{\mathcal{T}}$ and the full optimization provided by {\tt B} is confronting two very different algorithms, and shows that {\tt B} solves all of the full optimization problems that have been tested within milliseconds, whereas D-Wave's machine can sometimes take tenths of a second (before applying the multiplier factor $\simeq 2$, due to the binary search; see the appendix). When larger chips become available, however, it will be interesting to compare {\tt B} to a quantum annealing solver for sizes considered {\tt B}-intractable due to increasing memory and time requirements.

The comparison with the {\tt MS} method has a promising signature even now, with roughly half of the instances being solved by D-Wave's hardware faster than the {\tt MS} algorithm (with the caveat that our straightforward implementation is not fully optimized). Interestingly, the different parametrized ensembles of problems have distinctively different computational complexity characterized by well-recognizable average computational time to solution for {\tt MS} (i.e., the points are ``stacked around horizontal lines'' in \mbox{Figure \ref{fig:last}-b)}, whereas the D-Wave machine's complexity seems to be sensitive mostly to the total qubit count (see \mbox{Figure \ref{fig:last}-a)} irrespective of the problem class. We emphasize again that conclusions on speedup and asymptotic advantage still cannot be confirmed until improved hardware with more qubits and less noise becomes available for empirical testing.

\begin{figure}[t!]
\begin{centering}
\includegraphics[scale=0.6]{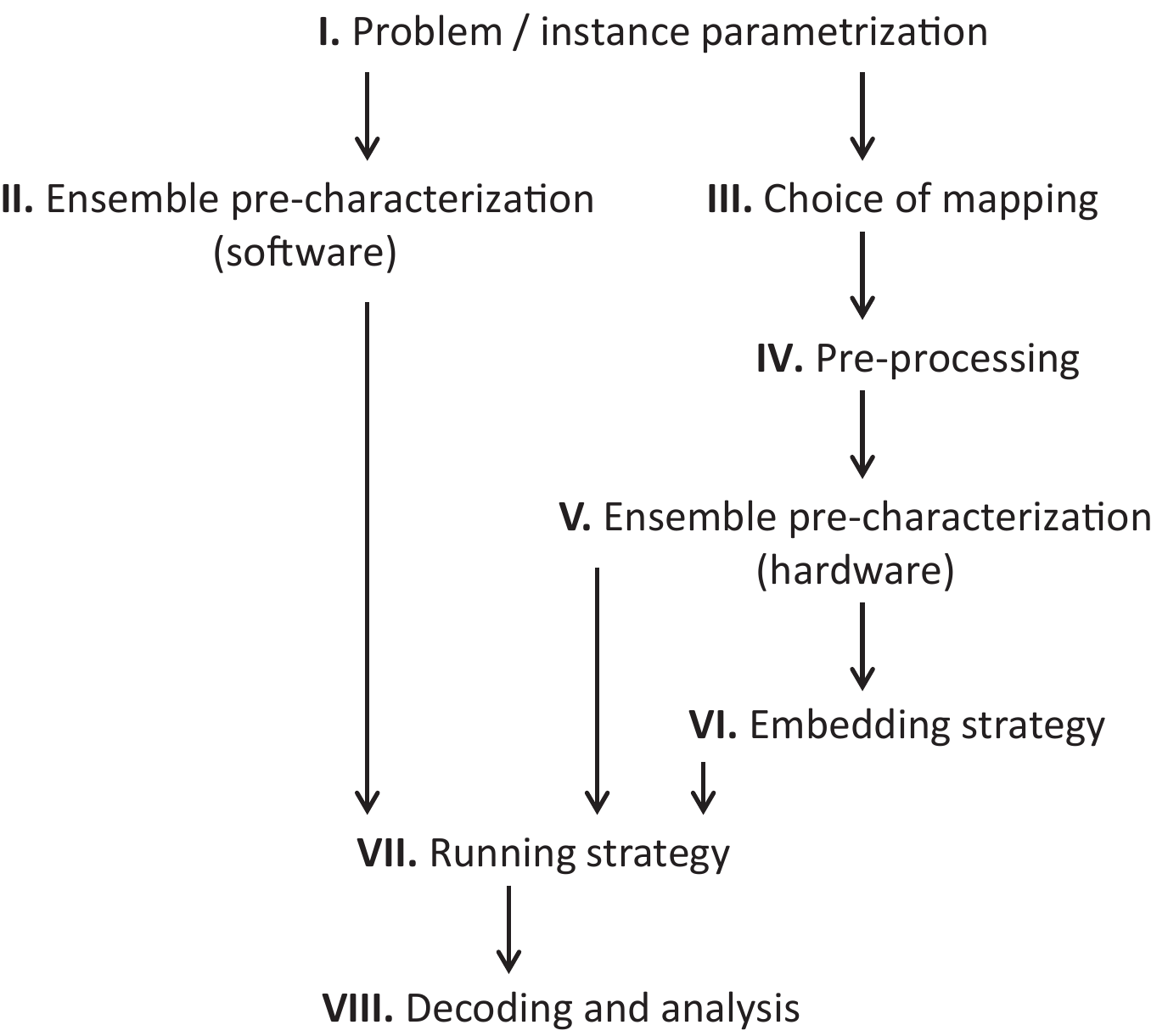}
\par\end{centering}

\caption{{I--II) Appropriate choice of benchmarking and classical simulations is discussed in Section IV. III--IV) Mapping to QUBO problems is discussed in Sections II and III. V--VI) Pre-characterization for parameter setting is described in Section VI. VII) Structured run strategies adapted to specific problems have not to our knowledge been discussed before. We discuss a prescription in the appendix. VIII) The only decoding required in our work is majority voting within embedding components to recover error-free logical solutions. The time-indexed formulation then provides QUBO problem solutions that can straightforwardly be represented as Gantt charts of the schedules.}}

\label{fig:summary}
\end{figure}

\section{VI. Conclusions}

Although it is probable that the quantum annealing-based JSP solver proposed herein will not prove competitive until the arrival of an annealer a few generations away, the implementation of a provably tough application from top to bottom was missing in the literature, and our work has led to noteworthy outcomes we expect will pave the way for more advanced applications of quantum annealing. Whereas part of the attraction of quantum annealing is the possibility of applying the method irrespective of the structure of the QUBO problem, we have shown how to design a quantum annealing solver, mindful of many of the peculiarities of the annealing hardware and the problem at hand, for improved performance. \mbox{Figure~\ref{fig:summary}} shows a schematic view of the streamlined solving process describing a full JSP optimization solver.
The pictured scheme is not intended to be complete, for example, the solving framework can benefit from other concepts such as performance  tuning techniques \cite{Perdomo2015} and error-correction repetition lattices \cite{vinci2015quantum}. The use of the decision version of the problem can be combined with a properly designed search strategy (the simplest being a binary search) in order to be able to seek the minimum value of the common deadline of feasible schedules.
The proposed timespan discrimination further provides an adjustable compromise between the full optimization and decision formulations of the problems, allowing for instant benefits from future improvements in precision without the need for a new formulation or additional binary variables to implement the makespan minimization as a term in the objective function. As will be explored further in future work, we found that instance pre-characterization performed to fine tune the solver parameters can also be used to improve the search strategy, and that it constitutes a tool whose use we expect to become common practice in problems amenable to CSP formulations as the ones proposed for the JSP. Additionally, we have shown that there is great potential in adapting classical algorithms with favorable polynomial scaling as pre-processing techniques to either prune variables or reduce the search space. Hybrid approaches and metaheuristics are already fruitful areas of research and ones that are likely to see promising developments with the advent of these new quantum heuristics algorithms. 

\subsection{Acknowledgements}

The authors would like to thank J. Frank, M. Do, E.G. Rieffel, B. O'Gorman, M. Bucyk, P. Haghnegahdar, and other researchers at QuAIL and 1QBit for useful input and discussions. This research was supported by 1QBit, Mitacs, NASA (Sponsor Award Number NNX12AK33A), and by the Office of the Director of National Intelligence (ODNI), Intelligence Advanced Research Projects Activity (IARPA), via IAA 145483 and by the AFRL Information Directorate under grant F4HBKC4162G001.

\appendix

\section{JSP and QUBO formulation}

In this appendix we expand on the penalty form used for the constraints and the alternative reward-base formulation as well as the timespan discrimination scheme.

\subsection{Penalties versus rewards formulation}\label{sec:rewards}

The encoding of constraints as terms in a QUBO problem can either reward the respecting of these constraints or penalize their violation. Although the distinction may at first seem artificial, the actual QUBO problem generated differs and can lead to different performance on an imperfect annealer. We present one such alternative formulation where the precedence constraint (\ref{eq:h1}) is instead encoded as a reward for correct ordering by replacing
$+\eta h_1(\bar{x})$
with
$-\eta' h'_1(\bar{x})$,
where
\begin{equation}
h'_1(\bar{x}) =
\sum_n\left(\sum_{\substack{k_{n-1} < i < k_n\\t+p_i\le t'}}x_{i,t}x_{i+1,t'}\right).
\end{equation}
The new Hamiltonian is
\begin{equation}
H'_T(\bar{x}) = -\eta' h'_1(\bar{x}) + \alpha h_2(\bar{x}) + \beta h_3(\bar{x}).
\end{equation}
The reward attributed to a solution is equal to $\eta'$ times the number of
satisfied precedence constraints. A feasible solution, where all
constraints are satisfied, will have energy equal to $-\eta'(k_N - N)$.

The functions  $h_1$ and $h'_1$ differ only by the range of $t'$: in the rewards
version we have  $$t' - t \ge p_i,$$ and in the penalties version we have
$$t'- t <  p_i.$$ The fact that we are allowing equality in
the rewards version means that $h'_1$ will always have more quadratic terms than $h_1$
regardless of variable pruning, leading to a more connected QUBO graph and 
therefore a harder problem to embed. 

Another important disadvantage is revealed when choosing the coefficients $\eta'$, $\alpha$, and $\beta$ in $H'_T$ to
guarantee that no unfeasible solution has energy less than $-\eta'(k_N - N)$.
This can happen if the penalty that we gain from breaking constraints
$h_2$ or $h_3$ is less than the potential reward we get from $h'_1$. The penalty-based formulation simply 
requires that  $\eta$, $\alpha$, and $\beta$ be larger than $0$. The following lemma summarizes the equivalent condition for the 
reward-based case.
\begin{lemma}
If $\beta/\eta' \geq 3$ and $\alpha > 0$, then
\begin{equation}\label{original_inequality}
H'_T(\bar{x}) \geq -(k_N - N),
\end{equation}
for all $\bar{x}$, with equality if and only if $\bar{x}$ represents a feasible
schedule.
\end{lemma}
We also found examples that show that these bounds on the coefficients
are tight.

The fact that $\beta/\eta'$ must be greater than or equal to 3 is a clear disadvantage
because of the issues with precision of the current hardware. In $H_T$ we can set
all of the penalty coefficients (and hence all non-zero couplers) to be equal,
which is the best possible case from the point of view of precision.

\subsection{Timespan discrimination}
 
The timespan discrimination that we propose is a specification to strike a compromise between the information obtained from each solver call, and the required precision for this information to be accurate and obtained efficiently. Specifically, we want this extra information to help identify the optimal makespan by looking at the energy of the solutions. This means breaking the degeneracy of the ground states (i.e., the valid solutions) and assigning different energy sectors to different makespans. To prevent collisions with invalid solutions, these energy sectors have to fit within the logical QUBO problem's gap given by $\Delta E=\min\{\eta, \alpha, \beta\}$. We note that this will affect the actual gap (as seen by the hardware) of the embedded Ising model. 

Since the binary variables we have defined in the proposed formulation are not sufficient to write a simple expression for the makespan of a given solution, additional auxiliary variables and associated constraints would need to be introduced. Instead, a simple way to implement this feature in our QUBO formulation is to add a number of local fields to the binary variables corresponding to the last operation of each job, $\{O_{k_1},\dots,O_{k_N}\}$. Since the makespan depends on the completion time of the last operation, the precedence constraint guaranties that the makespan of a valid solution will be equal to the completion time of one of those operations. We can then select the local field appropriately as a function of the time index $t$ to penalize a fixed number $K$ of the larger makespans ranging from $T-K+1$ to $T$. Within a sector assigned to the time step $\mathcal{T}$, we need to further divide $\Delta E_{\mathcal{T}}$ by the maximum number of operations that can complete at $\mathcal{T}$ to obtain the largest value we can use as the local field $h_\mathcal{T}$, i.e., the number of distinct machines used by at least one operation in the set of operations $\{O_{k_1},\dots,O_{k_N}\}$, denoted by $M_{\textrm{final}}$. If $K$ is larger than 1, we also need to ensure that contributions from various sectors can be differentiated. The objective is to assign a distinct $T$-dependent energy range to all valid schedules with makespans within $[T-K,\,T]$. More precisely, we relate the local fields for various sectors with the recursive relation
\begin{equation}
h_{\mathcal{T}-1}=\frac{h_\mathcal{T}}{M_{\textrm{final}}} +\epsilon,\label{eq:recursive_precision}
\end{equation}
where $\epsilon$ is the minimum logical energy resolvable by the annealer. Considering that $\epsilon$ is also the minimum local field we can use for $h_{\mathcal{T}-K+1}$ and that the maximum total penalty we can assign through this time-discrimination procedure is $\Delta E-\epsilon$, it is easy to see that the energy resolution should scale as $\Delta E / (M_{\textrm{final}}^{K})$.
An example of the use of local fields for timespan discrimination is shown in \mbox{Figure 1-d} of the main text for the case $K=1$.

\section{Computational strategy} 

This appendix elaborates on the compilation methods and the quantum annealing implementation of a full optimization solver based on the decision version and a binary search as outlined in the main text. 

\subsection{Compilation}

The process of compiling, or \emph{embedding}, an instance for a specific target architecture is a crucial step given the locality of the programmable interactions on current quantum annealer architectures. During the graph-minor topological embedding, each vertex of the problem graph is mapped to a subset of connected vertices, or subgraph, of the hardware graph. These assignments must be such that the the edges in the problem graph have at least one corresponding edge between the associated subgraphs in the hardware graph. Formally, the classical Hamiltonian Eq.~(\ref{eq:objective_function}) is mapped to a quantum annealing Ising Hamiltonian on the hardware graph using the set of equations that follows. The spin operators $s\vec{ \sigma}_i$ are defined by setting $s=1$ and using the Pauli matrices to write $\vec{\sigma}_i=(\sigma^x_i, \sigma^y_i, \sigma^z_i)$. The resulting spin variables $\sigma^z_i=\pm1$, our qubits, are easily converted to the usual binary variables $x_i=0,1$ with $\sigma_{i}^z=2 x_{i}-1$. The Ising Hamiltonian is given by    
\begin{eqnarray}
H &=& A(t) \left [ H_{Q}+H_{E} \right ]+ B(t) H_{D} , \\
H_{Q} &=&  \sum_{ij}  \ J_{ij}\sigma_{\alpha_i}^z \sigma_{\beta_j}^z \big|_{(\alpha_i,\beta_j)\in E(i,j)} +\!\!\!\!\sum_{i \atop k \in V(i)} \!\!\!\!\!\frac{h_i}{N_{V(i)}} \sigma_{k}^z ,\quad \\
H_{E} &=& - \!\!\!\!\!\sum_{i \atop (k,k^\prime) \in E(i,i)}  J_{i,k,k^\prime}^F \sigma_k^z \sigma_{k^{\prime}}^z , \label{eq:emb_hamilt_Jf} \\
H_{D} &=&  \sum_{i \atop k \in V(i)} \sigma_{k}^x , 
\end{eqnarray}
where for each logical variable index $i$ we have a corresponding ensemble of qubits given by the set of vertices $V(i)$ in the hardware graph with $|V(i)|= N_{V(i)}$. Edges between logical variables are denoted $E(i,j)$ and edges within the subgraph of $V(i)$ are denoted $E(i,i)$. The couplings $J_{ij}$ and local fields $h_{i}$ represent the logical terms obtained after applying the linear QUBO-Ising transformation to Eq.~(\ref{eq:objective_function}). $J_{i, k,k^{\prime}}^F$ are embedding parameters for vertex $V(i)$ and $(k,k')\in E(i,i)$ (see discussion below on the ferromagnetic coupling). The equation above assumes that a local field $h_i$ is distributed uniformly between the vertices $V(i)$ and the coupling $J_{i,j}$ is attributed to a single randomly selected edge $(\alpha_i, \beta_j)$ among the available couplers $E(i,j)$, but other distributions can be chosen. In the actual hardware implementation we rescale the Hamiltonian by dividing by $J_F$, which is the value assigned to all $J_{i,k,k^\prime}^F$, as explained below. This is due to the limited range of $J_{ij}$ and $h_i$ allowed by the machine~\cite{Venturelli2015SK}.

Once a valid embedding is chosen, the ferromagnetic interactions $J_{i, k,k^\prime}^F$ in Eq.~(\ref{eq:emb_hamilt_Jf}) need to be set appropriately. While the purpose of these couplings is to penalize states for which $\langle \sigma_{k}^z \rangle \neq \langle \sigma_{k^\prime}^z \rangle$ for $k,k^\prime \in V(i)$, setting them to a large value negatively affects the performance of the annealer due to the finite energy resolution of the machine (given that all parameters must later be rescaled to the actual limited parameter range of the solver) and the slowing down of the dynamics of the quantum system associated with the introduction of small energy gaps. There is guidance from research in physics~\cite{Venturelli2015SK, King2015} and mathematics~\cite{ChoiEmbedding2} on which values could represent the optimal $J_{i,k,k^\prime}^F$ settings, but for application problems it is customary to employ empirical prescriptions based on pre-characterization~\cite{RieffelVenturelli_qip} or estimation techniques of performance~\cite{Perdomo2015}.

Despite embedding being a time-consuming classical computational procedure, it is usually not considered part of the computation and its runtime is not measured in determining algorithmic complexity. This is because we can assume that for parametrized families of problems one could create and make available a portfolio of embeddings that are compatible with all instances belonging to a given family. The existence of a such a library would reduce the computational cost to a simple query in a lookup table, but this could come at the price of the available embedding not being fully optimized for the particular problem instance.

\subsection{Quantum annealing optimization solver} 

We now detail our approach to solving individual JSP instances. We shall assume the instance at hand can be identified as belonging to a pre-characterized family of instances for minimal computational cost. This can involve identifying $N$, $M$, and $\theta$, as well as the approximate distribution of execution times for the operations. The pre-characterization is assumed to include a statistical distribution of optimal makespans as well as the appropriate solver parameters ($J_{\textrm{F}}$, optimal annealing time, etc.). Using this information, we need to build an ensemble of queries $\mathcal{Q}=\{\mathtt{q}\}$ to be submitted to the D-Wave quantum annealer to solve a problem $H$. Each element of $\mathcal{Q}$ is a triple $(t_A,\,R,\,T)$ indicating that the query considers $R$ identical annealings of the embedded Hamiltonian $H_T$ for a single annealing time $t_A$. To determine the elements in $\mathcal{Q}$ we first make some assumptions, namely, i) {\it{sufficient statistics}}: for each query, $R$ is sufficiently large to sample appropriately the ensembles defined in Eqs. (\ref{eq:solver_a})--(\ref{eq:solver_c}); ii) {\it{generalized adiabaticity}}: $t_A$ is optimized (over the window of available annealing times) for the best annealing performance in finding a ground state of $H_T$ (i.e., the annealing procedure is such that the total expected annealing time $t_AR^\star$ required to evolve to a ground state is as small as possible compared to the time required to evolve to an excited state, with the same probability). Both of these conditions can be achieved in principle by measuring the appropriate optimal parameters $R^\star(\mathtt{q})$ and $t_A^\star(\mathtt{q})$ through extensive test runs over random ensembles of instances. However, we note that verifying these assumptions experimentally is currently beyond the operational limits of the D-Wave Two device since the optimal $t_A$ for generalized adiabaticity is expected to be smaller than the minimum programmable value~\cite{ronnow2014defining}. Furthermore, we deemed the considerable machine time required for such a large-scale study too onerous in the context of this initial foray. Fortunately, the first limitation is expected to be lifted with the next generation of chip, at which point nothing would prevent the proper determination of a family-specific choice of $R^\star$ and $t_A^\star$. Given a specified annealing time, the number of anneals is determined by specifying $r_0$, the target probability of success for queries or confidence level, and measuring $r_{\mathtt{q}}$, the rate of occurrence of the ground state per repetition for the following query:
\begin{equation}
R^\star = \frac{\log[1-r_0]}{\log [1-r_{\mathtt{q}}]}.\label{eq:Rstar}
\end{equation}
The rate $r_{\mathtt{q}}$ depends on the family, $T$, and the other parameters. The minimum observed during pre-characterization should be used to guarantee the ground state is found with at least the specified $r_0$. Formally, the estimated time to solution of a problem is then given by  
\begin{equation}
\mathtt{T}=\sum_\mathtt{q\in\mathcal{Q}} t_A \left( \frac{\log[1-r_0]}{\log [1-r_{\mathtt{q}}]} \right).
\end{equation}
The total probability of success of solving the problem in time $\mathtt{T}$ will consequently be $\prod_\mathtt{q}{r_\mathtt{q}}$. For the results presented in this paper, we used $R^\star=$ 500 000 and $t_A^\star=\min(t_A)=20$ $\mu$s.

We can define three different sets of qubit configurations that can be returned when the annealer is queried with $\mathtt{q}$. $\mathcal{E}$ is the set of configurations whose energy is larger than $\Delta E$ as defined in Section III of the paper. These configurations represent invalid schedules. $\mathcal{V}$ is the set of solutions with zero energy, i.e., the solutions whose makespan $\mathcal{T}$ is small enough ($\mathcal{T}\le T-K$) that they have not been discriminated by the procedure described in the subsection on timespan discrimination. Finally, $\mathcal{S}$ is the set of valid solutions that can be discriminated ($\mathcal{T}\in (T-K,T]$). Depending on what the timespan $T$ of the problem Hamiltonian $H_{T}$ and the optimal makespan $\mathcal{T}$ are, the quantum annealer samples the following configuration space (reporting $R$ samples per query):

\begin{align}
T<\mathcal{T}&\longrightarrow&\mathcal{V},\mathcal{S}={\emptyset}&\rightarrow E_0>\Delta E\label{eq:solver_a},\\
\mathcal{T}\in (T-K, T]&\longrightarrow&\mathcal{V}={\emptyset}&\rightarrow E_0\in(0,\Delta E]\label{eq:solver_b},\\
\mathcal{T}\le T-K&\longrightarrow&\mathcal{E},\mathcal{V},\mathcal{S}\neq{\emptyset}&\rightarrow E_0=0.\label{eq:solver_c} 
\end{align}

Condition (\ref{eq:solver_b}) is the desired case where the ground state of $H_T$ with energy $E_0$ corresponds to a valid schedule with the  optimal makespan we seek. The ground states corresponding to conditions (\ref{eq:solver_a}) and (\ref{eq:solver_c}) are instead, respectively, invalid schedules and valid schedules whose makespan could correspond to a global minimum (to be determined by subsequent calls). The above-described assumption ii) is essential to justify aborting the search when case (\ref{eq:solver_b}) is encountered. If $R$ and $t_A$ are appropriately chosen, the ground state will be preferentially found instead of all other solutions such that one can stop annealing reasonably soon (i.e., after a number of reads on the order of $R^\star$) in the absence of the appearance of a zero-energy solution. We can then declare this minimum-energy configuration, found within $(0,\Delta E]$, to be the ground state and the associated makespan and schedule to be the optimal solution of the optimization problem. On the other hand, we note that if $K=0$, a minimum of two calls are required to solve the problem to optimality, one to show that no valid solution is found for $T=\mathcal{T}-1$ and one to show that a zero-energy configuration is found for $T=\mathcal{T}$.
While for cases (\ref{eq:solver_b}) and (\ref{eq:solver_c}) the appearance of an energy less than or equal to $\Delta E$ heuristically determines the trigger that stops the annealing of $H_T$, for case (\ref{eq:solver_a}), we need to have a prescription, based on pre-characterization, on how long to anneal in order to be confident that $T<\mathcal{T}$. While optimizing these times is a research program on its own that requires extensive testing, we expect that the characteristic time for achieving condition (\ref{eq:solver_b}) when $T=\mathcal{T}$ will be on the same order of magnitude for this unknown runtime.

\subsection{Search strategy}

The final important component of the computational strategy consists in determining the sequence of timespans of the calls (i.e., the ensemble $\mathcal{Q}$). Here we propose to select the queries based on an optimized binary search that makes informed dichotomic decisions based on the pre-characterization of the distribution of optimal makespans of the parametrized ensembles as described in the previous sections. More specifically, the search is designed based on the assumption that the JSP instance at hand belongs to a family whose makespan distribution has a normal form with average makespan $\langle \mathcal{T} \rangle$ and variance $\sigma^2$. This fitted distribution is the same $\mathcal{P}_{p}$ described in Figure 2-a of the main text whose tails have been cut off at locations corresponding to an instance-dependent upper bound $T_{\textrm{max}}$ and strict lower bound $T_{\textrm{min}}$  (see the following section on bounds).

\begin{figure}[h]
\begin{centering}
\includegraphics[scale=0.38]{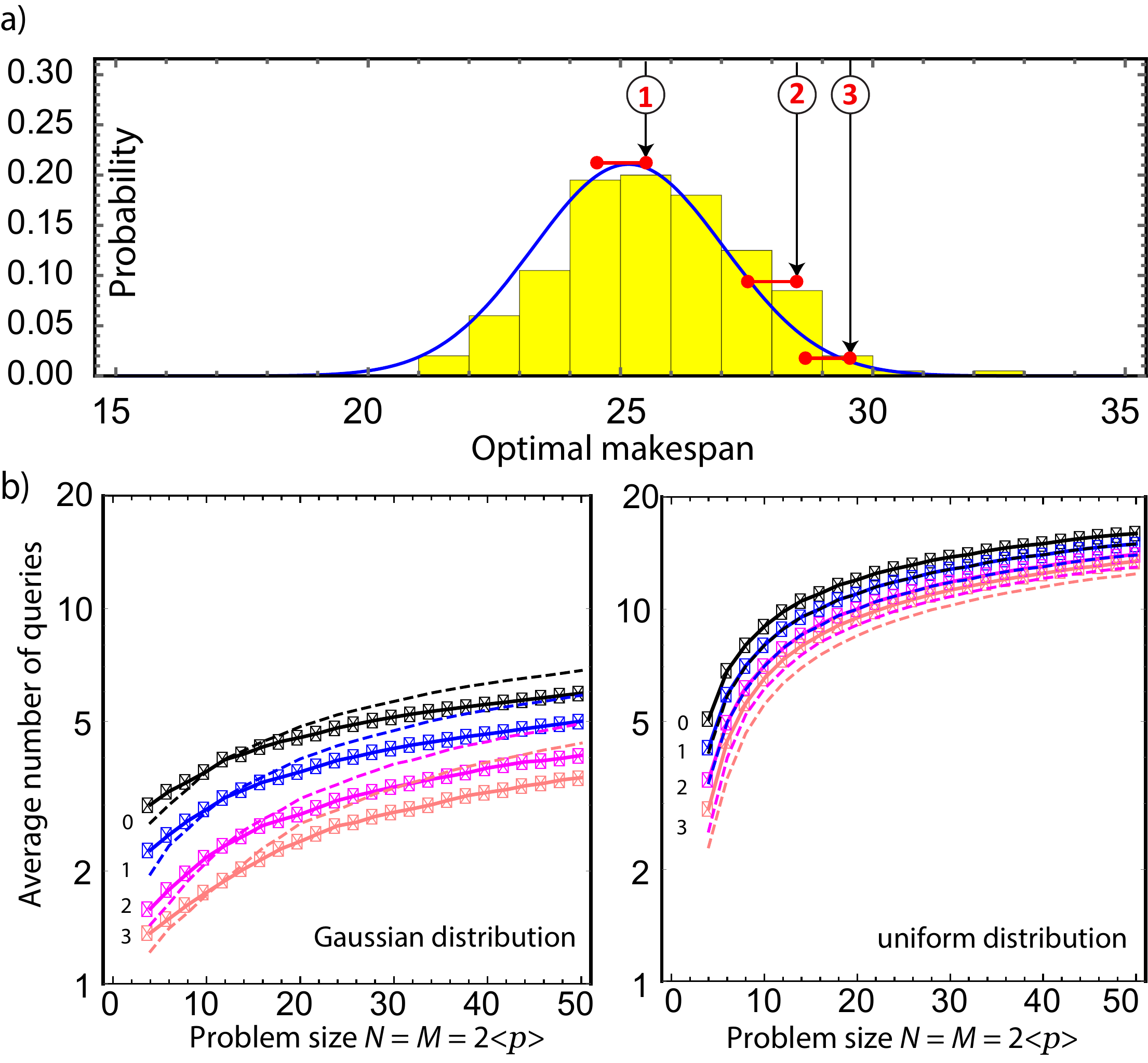}
\par\end{centering}

\caption{a) View of a guided binary search required to identify the minimum makespan over a distribution. The fitted example distribution corresponds to $N=M=8$, fitted to a Gaussian distribution as described in the main text. We assume $K=1$. The first attempt queries $H_{26}$, the second $H_{29}$, and the third $H_{30}$ (the final call), following Eq. (\ref{eq:binary}). b)~Average number of calls to the quantum annealer required by the binary search assuming Eq.~(\ref{eq:binary}) (left panel) or assuming a uniform distribution of minimum makespans between trivial upper and lower bounds. Thick and dashed lines correspond to $\theta=1$ and $\theta=0.5$, respectively, and the numeric values associated with each color in the figure correspond to different values of $K$. The operations' execution times are distributed uniformly with $P_{p}=\{0,\dots,N/2\}$.}

\label{fig:binary}
\end{figure}

Once the initial $T_{\textrm{min}}$ and $T_{\textrm{max}}$ are set, the binary search proceeds as follows. To ensure a logarithmic scaling for the search, we need to take into account the normal distribution of makespans by attempting to bisect the range $(T_{\textrm{min}},\,T_{\textrm{max}}]$ such that the probability of finding the optimal makespan on either side is roughly equal. In other words, $T$ should be selected by solving the following equation and rounding to the nearest integer: 
\begin{flalign}
\textrm{erf}&\bigg(\frac{T_{\textrm{max}}+\frac{1}{2}-\langle \mathcal{T} \rangle}{\sigma\sqrt{2}}\bigg)+\textrm{erf}\bigg(\frac{T_{\textrm{min}}+\frac{1}{2}-\langle \mathcal{T} \rangle}{\sigma\sqrt{2}}\bigg)= \label{eq:binary}\\
\textrm{erf}&\bigg(\frac{T+\frac{1}{2}-\langle \mathcal{T} \rangle}{\sigma\sqrt{2}}\bigg)+\textrm{erf}\bigg(\frac{T-\max(1,\,K)+\frac{1}{2}-\langle \mathcal{T} \rangle}{\sigma\sqrt{2}}\bigg)\nonumber,
\end{flalign}
where $\textrm{erf}(x)$ is the error function. For our current purpose, an inexpensive approximation of the error function is sufficient. In most cases this condition means initializing the search at $T=\langle \mathcal{T} \rangle$. We produce a query $\mathtt{q}_0$ for the annealing of $H_{T}$. If no schedule is found (condition (\ref{eq:solver_a})) we simply let $T_{\textrm{min}}=T$. If condition (\ref{eq:solver_c}) is verified, on the other hand, we update $T_{\textrm{max}}$ to be the makespan $\mathcal{T}$ of the valid found solution  (which is equal to $T-{\textrm{max}}(1,\,K)+1$ in the worst case) for the determination of the next query $\mathtt{q}_1$. The third condition (\ref{eq:solver_b}), only reachable if $K>0$, indicates both that the search can stop and the problem has been solved to optimality. The search proceeds in this manner by updating the bounds and bisecting the new range at each step and stops either with condition (\ref{eq:solver_b}) or when $\mathcal{T}=T_{\textrm{max}}=T_{\textrm{min}}+1$. Figure \ref{fig:binary}-a shows an example of such a binary search in practice. The reason for using this guided search is that the average number of calls to find the optimal makespan is dramatically reduced with respect to a linear search on the range $(T_{\textrm{min}},\,T_{\textrm{max}}]$. For a small variance this optimized search is equivalent to a linear search that starts near $T=\langle \mathcal{T} \rangle$. A more spread-out distribution, on the other hand, will see a clear advantage due to the logarithmic, instead of linear, scaling of the search. In Figure~\ref{fig:binary}-b, we compute this average number of calls as a function of $N$, $\theta$, and $K$ for $N=M$ instances generated  such that an operation's average execution time also scales with $N$. This last condition ensures that the variance of the makespan grows linearly with $N$ as well, ensuring that the logarithmic behavior becomes evident for larger instances. For this calculation we use the worst case when updating $T_{\textrm{max}}$ due to condition (\ref{eq:solver_c}) being met. We find that for the experimentally testable instances with the D-Wave Two device (see Figure 3-b of the main text), the expected number of calls to solve the problem is less than three (in the absence of pre-characterization it would be twice that), while for larger instances the size of $\mathcal{Q}$ scales logarithmically, as expected.

\subsection{JSP bounds}

The described binary search assumes that a lower bound $T_{\textrm{min}}$ and an upper bound $T_{\textrm{max}}$ are readily available. We cover their calculation for the sake of completeness. The simplest lower bounds are the \emph{job} bound and the \emph{machine} bound. The job bound is calculated by finding the total execution time of each job and keeping the largest one of them, put simply
\begin{equation}
\max_{n \in \{1,\,\dots,\, N\}} \sum_{i=k_{n-1}}^{k_n} p_i ,
\end{equation}
where we use the lexicographic index $i$ for operations and where $k_0=1$. 
Similarly, we can define the machine bound as
\begin{equation}
\max_{m\in\{1,\,\dots,\, M\}} \sum_{i\in I_m} p_i ,
\end{equation}
where $I_m$ is the set of indices of all operations that need to run on machine
${\bf m}_m$. Since these bounds are inexpensive to calculate, we can take the larger of the two. An even better lower bound can be obtained using the iterated Carlier and Pinson (ICP) procedure described in the window shaving subsection of Section III of the main text. We mentioned that the shaving procedure can
show that a timespan does not admit a solution if a window closes completely. Using shaving for different timespans and performing a binary search, we can obtain the ICP lower bound in $\mathcal{O}\big(N^2\log(N)M^2T_{\textrm{max}}\log_{2}(T_{\textrm{max}}-T_{\textrm{min}})\big)$, where $T_{\textrm{min}}$ and $T_{\textrm{max}}$ are some trivial lower and upper bound, respectively, such as the ones described in this section. Given that the search assumes a strict bound, we need to decrease whichever bound we chose here by one. 

As for the upper bound, an excellent choice is provided by another classical algorithm
developed by Applegate and Cook \cite{Applegate} for some finite computational effort. The
straightforward alternative is given by the total work of the problem
\begin{equation}
\sum_{i\in\{1,\,\dots,\, k_N\}} p_i. 
\end{equation}

The solver's limitations can also serve to establish practical bounds for the search. For a given family of problems, if instances of a specific size can only be embedded with some acceptable probability for timespans smaller than $T_{\textrm{max}}^\textrm{embed}$, the search can be set with this limit, and failure to find a solution will result in $T_{\textrm{max}}^\textrm{embed}$, at which point the solver will need to report a failure or switch to another classical approach. 


\section{Classical algorithms}

When designing a quantum annealing solver, a survey of classical methods provides much more than a benchmark for comparison and performance. Classical algorithms can sometimes be repurposed as useful pre-processing techniques as demonstrated with variable pruning. We provide a quick review of the classical methods we use for this work as well as some details on the classical solvers to which we compare.

\subsection{Variable pruning}\label{sec:shaving}

Eliminating superfluous variables can greatly help mitigate the constraints on the number 
of qubits available. Several elimination rules are available and we  
explain below in more detail the procedure we used for our tests.

The first step in reducing the processing windows is to eliminate unneeded variables by considering the precedence
constraints between the operations in a job, something we refer to as simple variable pruning.  
We define $r_i$  as the sum of the
execution times of all operations preceding operation $O_i$. Similarly, we
define $q_i$ as the sum of the execution times of all operations following $O_i$.
The numbers $r_i$ and $q_i$ are referred to as the \textit{head} and
\textit{tail} of operation $O_i$, respectively. An operation cannot start before
its head and must leave enough time after finishing to fit its tail, so the
window of possible start times, the \textit{processing window}, for operation $O_i$ is $[r_i, T -p_i -q_i]$.

If we consider the one-machine subproblems induced on each
machine separately, we can update the heads and tails of each operation and reduce
the processing windows further. For example, recalling that $I_j$ is the set of indices of
operations that have to run on machine $M_j$, we suppose that $a, b \in I_j$ are
such that
$$r_a + p_a + p_b + q_b > T.$$
Then $O_a$ must be run after $O_b$. This means that we can update $r_a$ with $$r_a =
\max\{r_a, r_b + p_b\}.$$ We can apply similar updates to the tails because of the
symmetry between heads and tails. These updates are known in the literature as
\emph{immediate selections}.

Better updates can be performed by using \emph{ascendant sets}, introduced by Carlier and Pinson in \cite{Carlier90}. A subset $X\subset I_j$ is
an ascendant set of $c\in I_j$ if $c \not\in I_j$ and
$$ \min_{a\in X\cup\{c\}}r_a + \sum_{a\in X\cup\{c\}}p_a + \min_{a\in X}q_a >
T.$$
If $X$ is an ascendant set of $c$, then we can update $r_c$ with
$$r_c = \max\left\{r_c, \max_{X'\subset X}\left[\min_{a\in X'}r_a + \sum_{a\in X'}p_j\right]\right\}.$$
Once again, the symmetry implies that similar updates can be applied to the tails.

Carlier and Pinson in \cite{Carlier} provide an algorithm to perform all of the ascendant-set
updates on $M_j$ in $\mathcal{O}(N\log(N))$, where $N = |I_j|$. After these
updates have been carried out on a per-machine basis, we propagate the new
heads and tails using the precedence of the operation by setting
\begin{equation}\label{propagate1}r_{i+1} = \max\left\{r_{i+1}, r_i +
p_i\right\},\end{equation} \begin{equation}\label{propagate2}q_i =
\max\left\{q_i, q_{i+1} + p_{i+1}\right\},\end{equation}
for every pair of operations $O_i$ and $O_{i+1}$ that belong to the same job.

After propagating the updates, we check again if  any ascendant-set updates can
be made, and repeat the cycle until  no more updates are found.  In our tests, we
use an implementation similar to the one described in \cite{Carlier} to do the
ascendant-set updates.

Algorithm \ref{pseudo} is pseudocode that describes the shaving procedure.
Here, the procedure UpdateMachine$(i)$ updates heads
and tails for machine $i$ in $\mathcal{O}(N\log(N))$, as described by Carlier and
Pinson in \cite{Carlier}. It returns True if any updates were made, and
False otherwise. \mbox{PropagateWindows()} is a procedure that iterates over the tasks
and checks that Eqs. (\ref{propagate1}) and (\ref{propagate2}) are satisfied,
in $\mathcal{O}(NM)$.

\begin{algorithm}
\caption{Shaving algorithm}\label{pseudo}
\begin{algorithmic}[1]
\Procedure{icp\_shave}{}
\State $updated \gets \text{True}$
\While {$updated$}
\State $updated \gets$ False
\For {$i \in $ machines}
\State $updated \gets \text{UpdateMachine}(i) \vee updated$
\EndFor
\If {$updated$}
PropagateWindows()
\EndIf
\EndWhile
\EndProcedure
\end{algorithmic}
\end{algorithm}

For each repetition of the outermost loop of Algorithm \ref{pseudo}, we know
that there is an update on the windows, which means that we have removed at
least one variable. Since there are at most $NMT$ variables, the loop will run at
most this many times. The internal \textbf{for} loop runs exactly $M$ times and does work
in $\mathcal{O}(N\log(N))$.
Putting all of this together, the final complexity of the shaving procedure is
$\mathcal{O}(N^2M^2T\log(N))$.

\subsection{Classical algorithm implementation}

Brucker et al.'s branch and bound method \cite{Brucker} remains widely used due to its state-of-the-art status on smaller JSP instances and its competitive performance on larger ones \cite{BruckerShave}. Furthermore, the original code is freely available through ORSEP \cite{BruckerCode}. No attempt was made at optimizing this code and changes were only made to properly interface with our own code and time the results.  

Martin and Shmoys' time-based approach \cite{Martin} is less clearly defined in the sense that no publicly available standard code could be found and because a number of variants for both the shaving and the branch and bound strategy are described in the paper. As covered in the section on shaving, we have chosen the $\mathcal{O}(n\log(n))$ variants of heads and tails adjustments, the most efficient choice available. On the other hand, we have restricted our branch and bound implementation to the simplest strategy proposed, where each node branches between scheduling the next available operation (an operation that was not yet assigned a starting time) immediately or delaying it. Although technically correct, the same schedule can sometimes appear in both branches because the search is not restricted to \emph{active} schedules, and unwarranted idle times are sometimes considered. According to Martin and Shmoys, the search strategy can be modified to prevent such occurrences, but these changes are only summarily described and we did not attempt to implement them. Other branching schemes are also proposed which we did not consider for this work. One should be careful when surveying the literature for runtimes of a full-optimization version based on this decision-version solver. What is usually reported assumes the use of a good upper bound such as the one provided by Applegate and Cook \cite{Applegate}. The runtime to obtain such bounds must be taken into account as well. It would be interesting to benchmark this decision solver in combination with our proposed optimized search, but this benchmarking we also leave for future work.

Benchmarking of classical methods was performed on an off-the-shelf Intel Core i7-930 processor clocked at 2.8~GHz.

\bibliography{draft_jsp_pra_sept}

\end{document}